# Performance analysis of SDN controllers: POX, Floodlight and Opendaylight

GERELTSETSEG Altangerel, TUGSJARGAL Chuluuntsetseg, DASHDORJ Yamkhin/Ph.D/

Department of Information network and Security
The school of Information and Communication Technology, Mongolian University of Science and Technology
Ulaanbaatar, Mongolia
gereltsetseg@must.edu.mn, b140970113@sict.edu.mn, dashdorj@must.edu.mn, en

*Abstract–*The IP network is time-consuming for configuration and troubleshooting because it requires access to every device command line interface (CLI) or Graphical User Interface (GUI). In other words, the control plane (gathering information to forward data) and data plane (data forwarding plane) are run on every intermediary device. To solve this problem, the software defined network emerged and separated the control plane (done by software) from the data plane (done by hardware) [1]. In addition, the control plane is operated in the central point named the controller. There are many controller software programs in simulation and production network environments. In this paper, we compare three open source controllers called POX, Floodlight and Opendaylight (ODL) in simulation network created by MININET SDN emulation program in terms of TCP and UDP throughput and average Round Trip Time (RTT) of the first packet of the flow in the mesh and tree topology.

*Keyword-SDN, OpenFlow, NOS(Network operating systems), Programmable network, Data plane, Control plane, SDN controllers, POX, Floodlight, Opendaylight*

## I. INTRODUCTION

As mentioned above, the traditional IP network has several disadvantages. Moreover, suppose ISP company will be installing new technologies such as IPv4 switching to IPv6. In this case, ISP engineers need access to the CLI and GUI of each device to introduce this new technology. Some ISPs have hundreds of devices while some others have thousands of devices. Furthermore, if devices come from different vendors such as Cisco, ZTE and Planet, the configuration commands and interfaces are different. Consequently, they have to work on those diverse interfaces.

Another example related to this problem is that it takes 5-10 years to introduce new developed routing protocols to the global network [2-3]. The main source of this problem is that the control plane and the data plane run together on every intermediary device. Furthermore, this disadvantage can lead to increased operating and labor budget. In order to address this limitation, the Software defined network (SDN) has been developed [4-5]. The main feature of SDN is that first the control plane is decoupled from device and second the control plane is done on controller. As a result, the intermediary device transmits the data according to the controller's instructions [6-7].

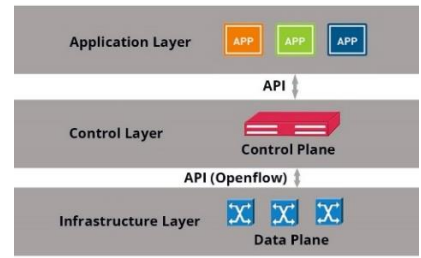

Fig 1. SDN network architecture

The Fig. 1 shows SDN network architecture with three layers. In order that these three layers communicate with each other, well-defined application programming interfaces (API) are developed. The interface between infrastructure layer and control layer is named Southbound API and the interface between the control layer and application layer is called northbound API. There are several southbound APIs such as Openflow, BGP-LS, LISP. But the OpenFlow developed by the Open Networking Foundation (ONF) is the first and probably most well-known southbound interface (API) [9-10]. The first version of Openflow 0.2.0 was released in 2008 and the last version (1.4.0) was released in 2013.

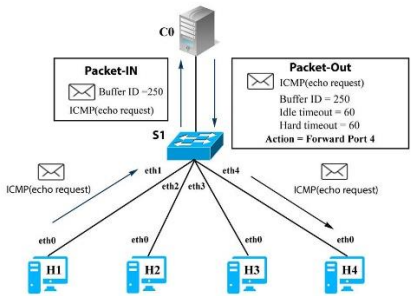

Fig 2. Packet handling example

How does the Openflow protocol work? The Fig. 2 shows the network with four hosts (H1, H2, H3, H4), a switch (S1) and a controller (C0). Let's consider that we send ICMP echo request from H1 (host) to H4 (host) through this network. Firstly, this packet will reach to the S1 (forwarding device). As the control plane is separated from S1, S1 doesn't know how to process this packet. Therefore, S1 will store this packet in its memory (Buffer-ID 250) and then will ask that how to process this ICMP echo request from C0 (controller) via the Openflow Packet-IN message. After C0 receiving this message, it will send Openflow Packet-Out message containing instructions to S1. According to this instruction, the flow table is created in switch (S1).

OpenFlow Packet-out message includes the following information:

- Buffer ID –Buffer ID is same with the Buffer ID in the Packet-IN message because this instruction is dedicated to process message stored in S1(buffer-id 250).
- Idle Timeout and Hard timeout – expresses how long this flow entry(instruction) is stored in S1.
- Action – this packet is forwarded to eth4.

Like this way flow table is created in data forwarding device like S1 in the Fig. 2.

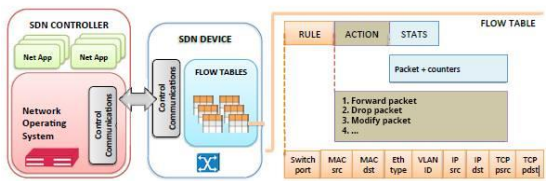

Fig 3. Flow structure

Data forwarding devices might have one or more flow tables. The Fig. 3 shows the flow structure. Based on the rule in the flow table, data may be forwarded, dropped or modified. Now it is obvious that these data processing rules in the flow table are determined by controller and transferred by OpenFlow protocol from the controller to the data forwarding devices. Based on what kind of rule is processed by controller software, data forwarding device acts as a router, switch, firewall or other device.

In recent years, OpenFlow and SDN technologies developed in academic environment have been used in practice. For instance, the vendors of network device are manufacturing devices that support OpenFlow API. Moreover, the Google, the Yahoo, the Facebook and the Microsoft are funded and worked together with Open Network Foundation (ONF), which developing open standards such as OpenFlow. Moreover, the Google has already implemented SDN in data center network. These are the proof of SDN's significance.

In this paper, we make performance analysis for open source and Openflow POX, ODL and Floodlight through average RTT time for the first packet of flow and TCP/UDP throughput in the tree and mesh topology by using MININET emulation program. Why we have chosen average RTT of the first packet of flow is that the first packet going to the SDN network is first processed as a controller. Based on the first packet processing rule, next packets are processed without connecting to the controller. Therefore, the first packet's responsive time (RTT) is critical. In some works relating to our research, average RTT of all packets of the flow was considered [14]. The main feature of our work is that we selected average RTT of the first packet of flow.

The paper is organized as follows: The second section pertains to the Software defined network, the third section on SDN controllers, the fourth section on Performance analysis of controller and finally the conclusions.

## II. SOFTWARE DEFINED NETWORK TECHNOLOGY

SDN technology initiated from research on Stanford University has four main pillars.

1. The control plane is separated from the data plane. By decoupling control plane (software) from network device, network became simple data forwarding node [11-12].
2. Sequence of packets with same source and destination address are called a flow. Data forwarding devices save state of the firstly transmitted packet of the flow how it is forwarded. The next packets of the flow are transmitted according to this state. In traditional network, data transfer process was based only on destination address in default whereas in SDN, it is based on multiple packet fields. Therefore it is not necessary to distinguish network devices as a router, switch and firewall [9].

3. The control plane is done on NOS (Network Operating system) or SDN controller. NOS is software to control data forwarding devices from single point. The function of this operating system is same with the traditional NOS. The difference is that this operating system runs only on the controller(not on all network devices).

4. The controller is directly connected to the data forwarding devices and programmed these devices [5].

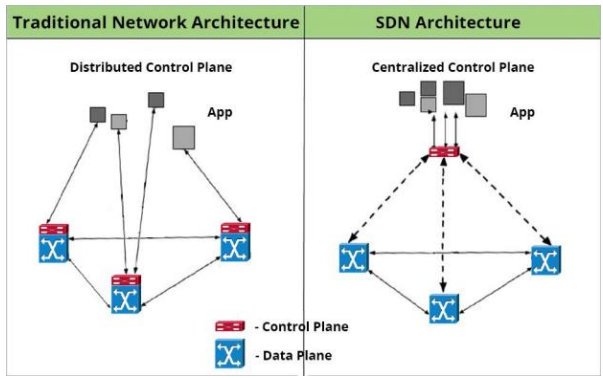

Fig 4. Traditional network versus SDN network

In the Fig. 4, whereas the control plane and the data plane are run on every intermediary device in traditional network, the control plane is done on controller in SDN. According to the controlling information of the controller, intermediary device forwards packets [13].

Furthermore, the vendor of network device produces both software and hardware in the traditional network. Therefore, customers can not reprogram their devices they want as the system is proprietary of the vendor. For instance, if you buy Nexus switch by Cisco, you can only use NX-OS accompanying with this hardware and it is not possible to change this OS through another operating system solution. But by implementing SDN technology, the controller can be programmed as you wish. It gives more flexibility than traditional network solution.

### III. SDN CONTROLLERS

SDN controller, managing network devices, is an application. There are so many controllers developed in various programming language. For example NOX, POX, Floodlight, Opendaylight, Beacon and so on.

Brief overview for some controllers.

- NOX: NOX, the first generation OpenFlow controller, is stable and widely used. Two versions are developed. The first one (NOX classic) is written in Python and the second one (NOX new) is in C++. The first one was shaded by the second one due to its performance [15].

- POX: POX is written in Python and Apache licensed. This is an easy-to-understand template for understanding SDN concepts.

- ODL: ODL, open source platform, is developed by the Linux Foundation in Java. Hydrogen, the first version, was released in 2014 whereas Fluorine, the last version, was released in 2018. Between these two, 7 releases are developed [16]. As ODL is open source, different projects and communities contribute its development. Some vendors such as the Big Switch and the Cisco also promoted ODL. It supports not only Openflow, but also other southbound APIs such as BGP-LS and LISP for added flexibility [15].

- Ryu: The main feature of Ryu, written in Python, has very large developer community. It supports the Openflow version 1.0-1.3 with Nicira extentions. It is also open source under the Apache 2.0 license.

- The Floodlight: It is an enterprise-class, Apache-licensed, Java-based OpenFlow controller. It is supported by the community of developers including a number of engineers from the Big Switch Networks. It is easy to extend and enhance because of its module loading system feature. Also, it supports virtual and physical switches [17].

If we list controllers like the above, it will cover several pages. We have selected three of the above controllers which are open source and OpenFlow controllers to make experiment. These controllers are also possible to be tested in MININET emulation program. The table 1 shows the main features of those controllers.

TABLE 1. MAIN FEATURES ON SELECTED CONTROLLERS

|  | POX(version) | Floodlight | Open daylight |
|---|---|---|---|
| Programming Language | Python | Java | Java |
| Northbound API | Open Flow | Open Flow | Open Flow, BGP-LS, LISP |
| Openflow version | 1 | 1.1 | 1.0-1.3 |
| License | Apache | Apache | Eclipse Public License (EPL) |
| Open source | Yes | Yes | Yes |
| Learning | Easy | Hard | Hard |
| GUI | No | Yes | Yes |
| Installation | Easy | Medium | Hard |

## IV. PERFORMANCE ANALYSIS ON CONTROLLERS

The experimental setup is shown in the Fig. 5. The first one is PC running MININET emulation program and the second one is Virtual Machine (VM). These two machines hardware and software specifications are in the Table 2. PC and VM are directly connected by the Ethernet LAN adapters (100Mbps).

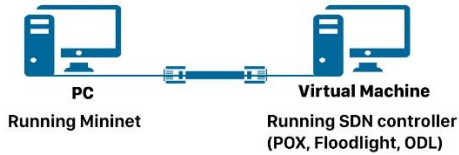

Fig 5. Experimental setup

TABLE 2. HARDWARE AND SOFTWARE SPECIFICTIONS

|  |  | PC | VM |
|---|---|---|---|
| Hardware | Processor | Intel(R) core i3-2100 | 2 core |
|  | RAM | 16 GB | 4 GB |
| Software | Operating System | Ubuntu 17.10 | Ubuntu 16.04.3 |
|  | Virtual Box | - | 5.2.8 |
|  | MININET | 2.3.0d1 | - |
|  | POX | - | 0.2.3 |
|  | ODL | - | 0.6.1-carbon |
|  | Floodlight | - | 1.2 |

The Fig. 6 and the Fig. 7 show the topology of the experimental setup for evaluating the performance of the POX, Floodlight and Opendaylight. Although the linear, single and tree built-in topologies are available in the Mininet, we have chosen only tree topology(Fig. 6) for our experiment because it is very popular in reality. Besides the tree topology, we have chosen the mesh topology which is programmed in Python by us. (Fig. 7).

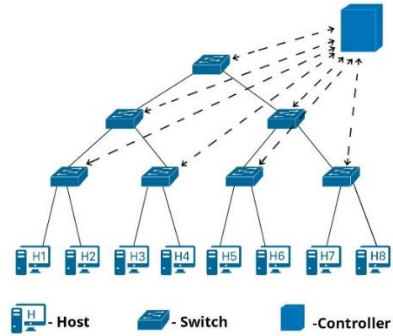

Fig 6. The tree topology in the Mininet

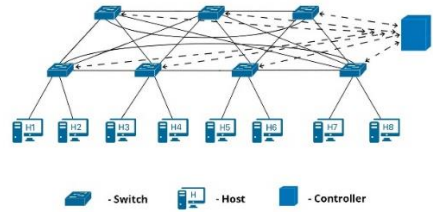

Fig 7. The mesh topology in the Mininet

From the above figures, each topologies (tree and mesh) have seven switches and eight hosts (H1-H8).

After all experimental setup was prepared, we began the first performance test. The first test is done on the tree topology with POX controller, the second on the tree topology with Floodlight, the third one on the tree topology with Opendaylight, the fourth on the mesh topology with POX, the fifth one on the mesh topology with Floodlight, and the last one on the mesh topology with Opendaylight. In each performance test, we send 50 packets with 1000 B and 3000 B size via ping command from host (H1) to host (H8), then we measured round trip time of the first packet of the flow for per test. These experiments and measurements are cycled 10 times for each test. Then we calculated the average values of measured RTTs of the first packet of the flow per test. These average values of the measured response times (RTT) in the first packet of flow are shown in the last two columns of the table 3 per test.

TABLE 3. TEST SETUP AND RTT FOR THE FIRST PACKET OF THE FLOW

|  | Topology | Controller | RTT(ms) for the first packet of the flow (Packet size is 1000B) | RTT(ms) for the first packet of the flow (Packet size is 3000B) |
|---|---|---|---|---|
| Test 1: | Tree topology | POX | 141 | 187 |
| Test 2: | Tree topology | Floodlight | 119 | 164 |
| Test 3: | Tree topology | ODL | 124 | 136.2 |
| Test 4: | Mesh topology | POX | 108.4 | 126.6 |
| Test 5: | Mesh topology | Floodlight | 34.1 | 45.1 |
| Test 6: | Mesh topology | ODL | 40.1 | 53.6 |

The Fig. 8 and 9 show comparison graph of average RTT time for the first packet of the flow for this each six experiment.

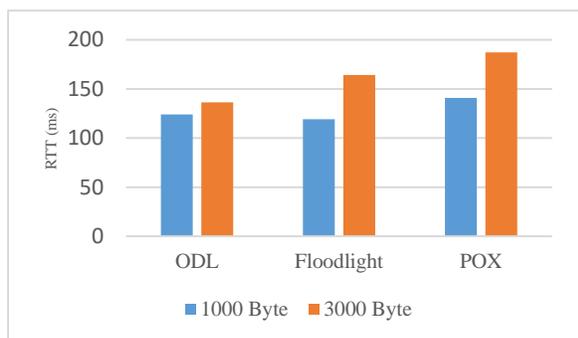

Fig 8. Average RTT for the first packet of the flow in the tree topology with different controllers

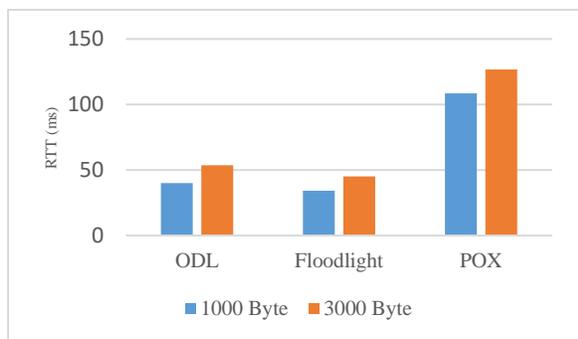

Fig 9. Average RTT for the first packet of the flow in the mesh topology with different controllers

From the above figures, average RTT for the first packet of the flow is the maximum for POX controller in both topologies. For tree topology, ODL and Floodlight controllers' performance in terms of average RTT for the first packet of the flow is almost same. In the mesh topology, average RTT for the first packet of the flow is the minimum for Floodlight controller. Also, difference between average RTT for the first packet of the flow in the mesh topology is greater among all other controllers. It is related to Spanning Tree Protocol(STP), which is running the mesh topology to prevent loop.

In the second performance test, we measured TCP and UDP flow throughput via iperf command for each six test. H1 acts as a TCP server and H8 works as a TCP client. From TCP client to TCP server, we sent packets for 20 seconds via 3 parallel connection and then we measured throughput for each simulation. Then H1 acts as a UDP server and H8 works as a UDP client. From UDP client to UDP server, we sent packets for 20 seconds via three parallel connection and then we measured throughput for each simulation. These attempts have been repeated 10 times and the average throughput of TCP and UDP was calculated for each 10 repeated test. The Fig. 10, 11, 12 and 13 are shown the graphical representation of the average TCP and UDP throughput of all our test.

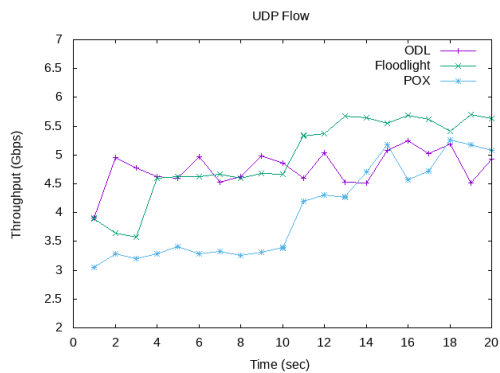

Fig 10. UDP flow throughput in the tree topology with different controllers

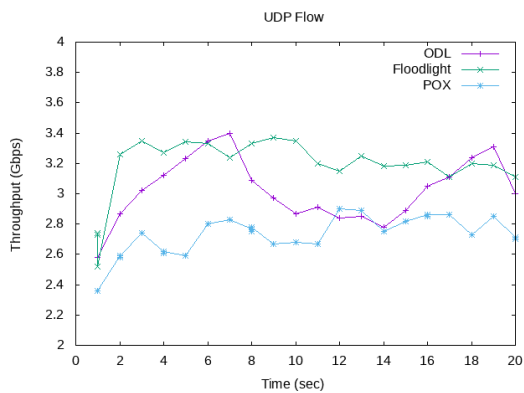

Fig 11. UDP flow throughput in the mesh topology with different controllers

From the above figures of UDP flow for the mesh and the tree topology, throughput of POX controller is minimum, whereas throughput of Floodlight controller is maximum for both of the topology. For ODL controller, the throughput is more fluctuated compared with other two controllers in the mesh and tree topology.

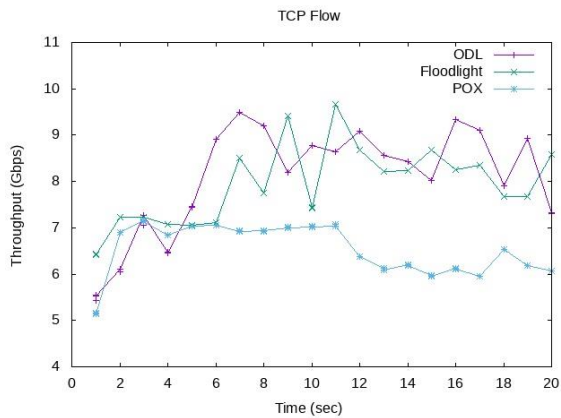

Fig 12. TCP flow throughput in the tree topology with different controllers

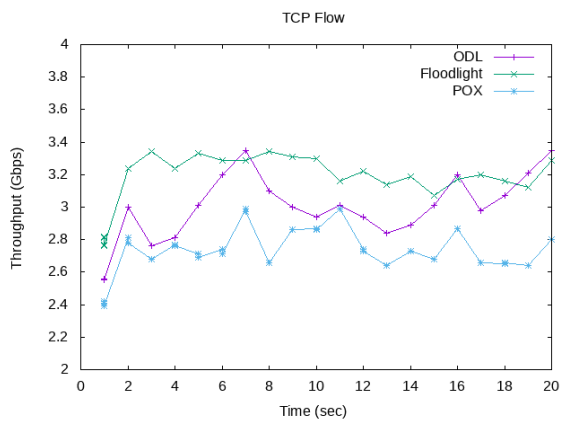

Fig 13. TCP flow throughput in the mesh topology with different controllers

POX controller's throughput is also minimum for TCP flow like UDP flow for both of the topology. Yet, throughput of Floodlight and ODL is near. Throughput fluctuation is slight for POX controller in tree topology, whereas throughput fluctuation is more for all controllers in mesh topology.

## III. CONCLUSION

SDN is new paradigm for fixing traditional network architectural limitation. All data processing decisions are made by controller. That is like the brain of network. There are plenty of controller software. In this paper, we made performance analysis on the three open source, OpenFlow controllers POX, Floodlight and ODL based on average round trip time for the first packet of the flow, TCP and UDP flow throughput. As a result of our experiment, Floodlight and ODL is better than POX controller in terms of packet response time (RTT) and throughput.

GERELTSETSEG Altangerel has been working as a lecturer at the Department of Information network, and security, the School of Information and Communication Technology in MUST. She is received his Bachelor in Information network in 2008 and his MSc in Information network from Mongolian University of Science and Technology in 2010 respectively. Her main research interests include network performance optimization, core network technologies and future internet.

TUGSJARGAL Chuluuntsetseg is studying as a system security major at the Department of Information network and security, the School of Information and Communication Technology in MUST to earn bachelor degree. Her main research interests include SDN, virtualization and network traffic analysis.

DASHDORJ Yamkhin is the head of Department of Information network and security, the School of Information and Communication Technology in MUST. He is received his Bachelor in Computer engineering at Novosibirsk State University, Russia in 1991, his MSc in Computer science from Mongolian University of Science and Technology and his PhD degree in Computer Science from Hanyang University of South Korea in 2009 respectively. He worked at Network center of MUST as a network administrator. His main research interests include network traffic analysis, modelling and embedded system.